\documentclass[11pt]{article}
\usepackage{amssymb,color}
\newcommand{\N}{N\raise.7ex\hbox{\underline{$\circ $}}$\;$}

\begin{document}

\textwidth 170mm \textheight 233mm \voffset -20mm \hoffset -15mm

\title {Hydrogen atom in de Sitter spaces}

\maketitle

\author{O.V. Veko\footnote{Kalinkovichi Gymnasium,
Belarus,vekoolga@mail.ru},
K.V. Kazmerchuk\footnote{Mosyr State
Pedagogical University, Belarus, kristinash2@mail.ru},
E.M. Ovsiyuk\footnote{Mosyr State Pedagogical University, Belarus,
e.ovsiyuk@mail.ru},
V.M. Red'kov\footnote{B.I. Stepanov Institute
of Physics, NAS of Belarus, redkov@dragon.bas-net.by},
A.M. Ishkhanyan\footnote{ Institute for Physical Research, Armenian
Academy of Sciences, aishkhanyan@gmail.com}
}

\begin{abstract}
The hydrogen atom theory is developed for the de
 Sitter and anti de Sitter spaces  on the basis of the Klein-Gordon-Fock wave equation in static coordinates. In both models, after separation of the variables, the problem is reduced to the general Heun equation, a second order linear differential equation having four regular singular points. A qualitative examination shows that the energy spectrum for the hydrogen atom in the de Sitter space should be quasi-stationary, and the atom should be unstable. We derive an approximate expression for energy levels within the quasi-classical approach and estimate the probability of decay of the atom. A similar analysis shows that in the anti de Sitter model the hydrogen atom should be stable in the quantum-mechanical sense. Using the quasi-classical approach, we derive approximate formulas for energy levels for this case as well. Finally, we present the extension to the case of a spin 1/2 particle for both de Sitter models. This extension leads to complicated  differential equations with 8 singular points.

\end{abstract}




\section{ Introduction}

Starting from the pioneering work by Dirac \cite{Dirac-1935}, a steady attention is
devoted to the de Sitter geometrical models in the context of quantum theory for curved
space-time (see, for instance, \cite{Hawking-Ellis-1973, Gibbons-2000}).
In particular, the problem of description of the particles with different spins on such
curved backgrounds has a long history (see \cite{Dirac-1935}--\cite{Otchik-Red'kov}). The most of the work done, however, concerns the field theory on the background of such geometrical models. In the present paper, we address the quantum mechanics and examine the influence of the de Sitter geometries on the behavior of the hydrogen atom. Since for the de Sitter spaces, when they
are  parameterized by static coordinates, no nonrelativistic Schr\"{o}diner equation
 is known, we have to employ the relativistic quantum mechanical approach,
 the Klein-Gordon-Fock scalar equation and the Dirac spinor equation. We consider these equations for both \mbox{de Sitter} ($dS$) and anti de Sitter ($AdS$) geometries.

In $dS$ space, we first analyze the expression for the classical
squared radial momentum $p_r^{2}(r)$.
We show that the equation $p_r^{2}=0$ may
have three positive and one negative roots, which correspond to a
particle moving in a potential well with a domain forbidden for
classical motion and a domain permitted for classical motion. In
other words, in the de Sitter geometry the hydrogen atom should be
unstable in the quantum-mechanical sense. The corresponding
 radial equation is reduced to the general Heun
equation having four regular singular points \cite{Ronveaux},
\cite{Slavyanov-Lay}. Studying the solutions of this equation, we
are led to a fourth order polynomial equation the roots of which
allow approximate calculation of the energy levels using the
WKB-method. The probability of decay of the atom is then
estimated.

A similar treatment is presented for the case of the anti de Sitter model.
The classical equation $p^{2}_{r}=0$ is again reduced to a fourth order polynomial
equation, which, however, this time may have two positive and two negative roots.
This configuration corresponds to a particle moving in a perfect potential well
 without tunneling. In other words, this means that in the $AdS$-model
  the hydrogen atom is a stable quantum-mechanical system. The corresponding radial equation is again reduced to the general Heun equation having four singular points. Discussing the solutions of this equation, we again consider a relevant fourth order polynomial equation, the roots of which provide approximate expressions for energy levels using the  WKB-method.

When treating the hydrogen atom using the Dirac equation, the radial differential equations arising after separation of the variables are much more complicated compared with the ones given by the scalar theory. Here, the problem is reduced to a differential equation with 8 singular points (note that the Dirac equation with Coulomb potential in hyperbolic Lobachevsky and spherical Riemamnn models is reduced to differential equations with 5 singular points \cite{Book-2012}). However, the physical
situation here is expected to be qualitatively similar to that arising
 in scalar models, because the influence of any special space-time geometry
 is universal and, in principle, it should not depend on the spin of
 a quantum-mechanical particle. The dependence on the value of the spin should  reveal only technical differences.

\section{Separation of the variables in $dS$ space}

If the origin of the coordinate system coincides with the location of the point electric charge, the Maxwell equations provide the 4-potential $A_{\alpha} = ( e / r, 0,0,0)$ for the Coulomb field.
We consider the  conventional generally covariant Klein-Gordon-Fock equation
\begin{eqnarray}
\left [\; \left (i\hbar \nabla _{\alpha} + {e \over c} A_{\alpha}
\right )
 \left  (i\hbar \nabla ^{\alpha} + {e \over c} A^{\alpha} \right ) - M^{2}c^{2} \right  ]
\Phi = 0 \; \label{2.2}
\end{eqnarray}

\noindent
in the de Sitter space model given by the static metrics \cite{Hawking-Ellis-1973}
\begin{eqnarray}
dS^{2} = \left (1- {r^{2} \over \rho^{2}} \right ) dt^{2}  - \left
(1- {r^{2} \over \rho^{2}} \right )^{-1} dr^{2} - r^{2}\left  ( d
\theta^{2} + \sin^{2} \theta d \phi^{2}\right  ) , \quad r\in [0,\rho) \; .
\label{2.1}
\end{eqnarray}
Note that in these coordinates no nonrelativistic Schr\"{o}dinger-type equation is known, so only relativistic description based on either scalar Klein-Gordon-Fock equation or spinor Dirac equation is possible.

After separation of the variables with the help of the substitution
$ \Phi = e^{-i\epsilon t / \hbar } Y_{lm}(\theta ,\phi ) f(r) $,
we arrive at the radial equation
\begin{eqnarray}
{ d^{2} \over d r^{2} } \; f\;  +        \; {2 (1-2r^{2} /
\rho^{2}) \over r (1- r^{2} / \rho^{2}) } \; {d \over dr } \; f\; + \qquad \qquad
\nonumber
\\
\left [ { ( \epsilon + e^{2} / r ) ^{2} \over c^{2} \hbar ^{2} }
{ 1 \over (1 - r^{2} / \rho^{2} )^{2} } - \left ( { M^{2} c^{2}
\over \hbar^{2} } + { l(l+1) \over r^{2} }\right ) {1 \over 1 -
r^{2} / \rho^{2} }\right  ] f = 0 \;. \label{2.3}
\end{eqnarray}

The Hamilton-Jakobi equation standing for the corresponding classical problem reads
\begin{eqnarray}
g^{00} \left ( {\partial S \over \partial x^{0}}   - {e \over c}
A_{0} \right )^{2} + g^{rr} \left ({\partial S \over \partial r}
\right )^{2} + g^{\theta \theta} \left ({\partial S \over \partial
\theta } \right )^{2} + g^{\phi \phi } \left ({ \partial S \over
\partial \phi } \right)^ {2} -M^{2} c^{2} =0 \; . \label{2.4}
\end{eqnarray}

\noindent Since the trajectory should be flat, one can fix the
coordinate  $\theta = {\pi/2}$, and the action function is
to be searched in the form
\begin{eqnarray} S(t,r,\phi) = -
\epsilon \;t + L \phi + \int p(r) dr \;, \quad L = {\partial S \over \partial \phi }\;  .
\label{action}
\end{eqnarray}
Eq. (\ref{2.4}) then gives the following expression for the  squared radial momentum:
\begin{eqnarray}
p^{2}_{r} =  { (\epsilon+ e^{2} / r) ^{2}  \over c^{2}  } \; {1
\over (1 - r^{2} / \rho^{2})^{2} } -  \left (M^{2} c^{2} +  { L^{2} \over
r^{2} } \right ) \; {1 \over 1 - r^{2} /\rho^{2} } \; . \label{2.5}
\end{eqnarray}

\noindent
Thus, the classical problem of a particle in the Coulomb field is reduced to the integral\footnote{This integral is calculated in elliptic functions.}
$\int p_{r} dr,$ where  $p_{r}$ is given by Eq. (\ref{2.5}), whereas the quantum-mechanical problem is reduced to the differential equation (\ref{2.3}).

\section{Qualitative discussion}

To study the expression (\ref{2.5}) for the squared radial momentum, it is convenient to compare with the flat Minkowski case, for which the squared momentum reads
\begin{eqnarray}
p^{2}_{r} = {(\epsilon + e^{2} / r )^{2} \over c^{2} } - \left ( M^{2}
c^{2} +
 {L^{2} \over r^{2}}\right ) .
\label{3.3a}
\end{eqnarray}

Consider first the free particle motion.
In the Minkowski space $p^{2}_{r}$ vanishes at two points:
\begin{eqnarray}
r_{0} = \pm \sqrt{L^{2}c^{2} \over \epsilon^{2} -
M^{2}c^{4} } \; . \label{3.1}
\end{eqnarray}

\noindent Accordingly, the classical motion is possible in the domain where
$p_{r}^{2} >0$, and at $\epsilon > Mc^{2}$ such a domain exists.

For a free particle in $dS$-space, at the origin and at the
horizon, the momentum  (\ref{2.5}) behaves as
\begin{eqnarray}
p_{r}^{2}(r \rightarrow 0 ) \sim - {L^{2} \over  r^{2} } \rightarrow -\infty \; , \qquad p_{r}^{2}(r
\rightarrow \rho )
 \; \sim \; { \epsilon^{2}  \over  c^{2} (1 - {r^{2} \over \rho^{2}})^{2}}
 \rightarrow +\infty \; .
\label{beh-1}
\end{eqnarray}

\noindent We find the vanishing points of momentum from the fourth order polynomial equation $p^{2}_{r}=0$:
\begin{eqnarray}
r_{0} = \pm \rho \sqrt{ -{A\over 2} \pm \sqrt{{A^{2} \over 4} +
{L^{2} \over M^{2} c^{2} \rho^{2}}}} \; , \;\;   \;\;\; A =
{\epsilon^{2}  - M^{2} c^{4} \over M^{2} c^{4} } + { L^{2} \over
M^{2} c^{2} \rho^{2} } \; .  \label{3.2}
\end{eqnarray}

\noindent Two of these roots are complex conjugate. Two others are real-valued, one of which is negative and another one is positive. We conclude that the character of the classical motion  of a free particle in the de Sitter space is much the same as in the flat Minkowski space.

Now, consider the case of the Coulomb potential.
For the particle on the background of the flat Minkowski space
the behavior of the momentum near the singular points is
\begin{eqnarray}
p_{r}^{2} (r  \rightarrow 0 ) \sim - {L^{2} - e^{4} / c^{2} \over r^{2} } \rightarrow -\infty
\; , \qquad    p_{r}^{2} (r  \rightarrow \infty ) \sim {\epsilon^{2} -
M^{2} c^{4} \over  c^{2} }=\textrm{const} .
\label{beh-2}
\end{eqnarray}

\noindent The turning points are given as
\begin{eqnarray}
r_{0} =  { c^{2} \over \epsilon^{2} - M^{2}c^{4}} \;  \left [ \; -{e^{2}
\epsilon \over c^{2}} \pm \sqrt{{e^{4}\epsilon^{2} \over c^{4}} +
\left(L^{2} - {e^{4}\over c^{2}}\right) {\epsilon^{2} - M^{2} c^{2} \over
c^{2}}} \;  \; \right ] . \label{3.3b}
\end{eqnarray}

\noindent If $(L^{2} - e^{4}/c^{2}) > 0$ and $\epsilon < Mc^{2} $,
with an additional requirement for the term under the square root to be positive,
according to (\ref{3.3b}) we have two positive roots and two negative (nonphysical) ones. The plot of $p_{r}^{2}$ is shown schematically in Fig. 1.

\vspace{10mm}

\unitlength=0.5mm
\begin{picture}(100,80)(-60,-40)
\special{em:linewidth 0.4pt} \linethickness{0.4pt}

\put(0,0){\vector(+1,0){150}}   \put(150,-5){$r$}
\put(0,-40){\vector(0,+1){80}}  \put(+2,+40){$p_{r}^{2}$}

\put(+10,0){\line(-1,-4){5}} \put(+5,-40){\line(0,+1){20}}

\put(+10,0){\line(+1,+1){20}} \put(+30,+20){\line(+1,-1){20}}

\put(+50,+0){\line(+2,-1){20}} \put(+70,-10){\line(+1,0){60}}

\end{picture}

\begin{center}

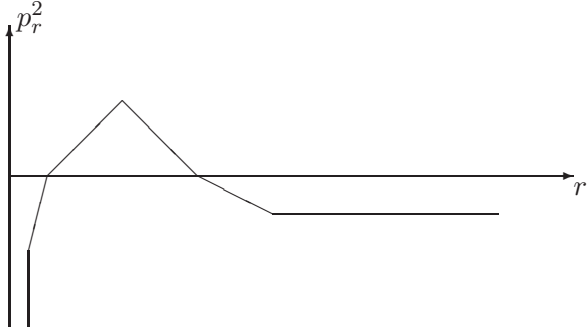
{\bf Fig. 1. Coulomb problem in the Minkowski space.}

\end{center}

\vspace{5mm}

For the de Sitter space, at the origin and at the horizon the momentum $p_{r}^{2}$ behaves as
\begin{eqnarray}
p_{r}^{2} (r \rightarrow 0)  \sim - { L^{2} - e^{4} /c^{2} \over
r^{2} }  \rightarrow - \infty  \; , \qquad
p_{r}^{2} (r \rightarrow \rho)  \sim  {(\epsilon +e^{2} / \rho )^{2}  \over  c^{2}\left  (1 -{r^{2}
/\rho^{2}} \right )^{2}}
 \rightarrow + \infty   \; .
\label{3.4b}
\end{eqnarray}
\noindent Furthermore, by Vieta's theorem, for the product of the roots $r_{1},...,r_{4}$
of the equation $p_{r}^{2}=0$ holds
\begin{eqnarray}
r_{1}r_{2}r_{3}r_{4}= - {(L^{2} - e^{4} /c^{2})\rho^{2}
\over {M^{2}c^{2}}}<0 \; .
\label{r1r2r3r4}
\end{eqnarray}

\noindent From here, we conclude that the following three variants
for the signs of the roots:
\begin{eqnarray}
\textrm{sign}(r_{1}, r_{2}, r_{3}, r_{4}) = (-,-,+, +)\; ,\; (-,-,-, -)\; ,\; (+,+,+, +)\; , \label{4.11a}
\end{eqnarray}

\noindent are excluded.
In turn, assuming two real-valued and two conjugate roots, we get
the variant
 $ \textrm{sign}(r_{1}, r_{2}, r_{3}, r_{4})
   = (-,+, z, z^{*})$,
 which  has no physical sense. Hence, the relationship
$
r_{1}r_{2}r_{3}r_{4} < 0$
 shows that, assuming all four roots are real, there are only two possibilities:
\begin{eqnarray}
\textrm{sign}(r_{1}, r_{2}, r_{3}, r_{4}) = (-,-,-,+)\; ,\; (-,+,+,+)\; . \label{4.11b}
\end{eqnarray}

\noindent  By physical arguments, more interesting is the
second case. The schematic behavior of $p_{r}^{2}$ corresponding
to this case is shown in Fig. 2. This behavior indicates that we should have
a nonstable quantum-mechanical system.

\vspace{10mm}

\unitlength=0.5mm
\begin{picture}(100,80)(-80,-40)
\special{em:linewidth 0.4pt} \linethickness{0.4pt}

\put(0,0){\vector(+1,0){130}}   \put(130,-5){$r$}
\put(0,-40){\vector(0,+1){80}}  \put(+2,+40){$p_{r}^{2}$}

\put(+115,-5){$\rho$}   \put(+115,-0){\circle*{2}}

\put(+10,0){\line(-1,-4){5}} \put(+5,-40){\line(0,+1){20}}

\put(+10,0){\line(+1,+1){20}} \put(+30,+20){\line(+1,-1){20}}

\put(+50,+0){\line(+2,-1){20}} \put(+70,-10){\line(+1,0){20}}
\put(+90,-10){\line(+1,0){10}} \put(+100,-10){\line(+1,+1){10}}

\put(+110,0){\line(+1,+4){5}} \put(+115,+20){\line(0,+1){20}}

\put(+5,+2){$r_{1}$} \put(+52,+2){$r_{2}$} \put(+105,+2){$r_{3}$}

\end{picture}

\begin{center}

{\bf Fig. 2. Coulomb problem in $dS$ space.}

\end{center}

\vspace{10mm}

Let us discuss in more detail the behavior of $p_{r}$ at the horizon $r
\rightarrow \rho $. To do this, we consider the radial velocity measured at the proper time:
\begin{eqnarray}
p_{r} = Mc {d r \over d S} \; , \;\; d S = \left(1 - {r^{2} \over
\rho^{2}}\right) c d \tau \; , \qquad  V^{2}_{r} = \left({dr \over d \tau
}\right)^{2} = {p^{2} _{r} \over M^{2} c^{2}} \left(1 - {r^{2} \over
\rho^{2}}\right)^{2} \; , \label{3.5a}
\end{eqnarray}

\noindent that is
\begin{eqnarray}
V_{r} = c \sqrt{{(\epsilon + e^{2} / r)^{2} \over M^{2} c^{4} } -
\left ( 1 + {L^{2} \over M^{2} c^{2}r^{2}} \right )\left  (1 - {r^{2} \over
\rho^{2}} \right )}  \; . \label{3.5b}
\end{eqnarray}

\noindent Correspondingly, at the horizon we have
\begin{eqnarray}
V_{r} = c { \sqrt{(\epsilon + e^{2} / \rho)^{2} } \over Mc^{2} }
\;\; \;\Longrightarrow \;\; \; \left\vert \epsilon + {e^{2} \over \rho } \right\vert
< M c^{2} \; , \label{3.5c}
\end{eqnarray}

\noindent since we assume that $V_{r}$ cannot be greater than the speed of the light $c$.

There is another argumentation, which is more appropriate in the context of quantum-mechanics.
Indeed, because the coordinate $r$ involved in the wave equation has no direct metrical sense,
instead of it one can use any other coordinate. To find such a  more convenient
variable, we require the radial momentum in that new variable to be finite
at $r \rightarrow \rho$. However, $p_{r} = (\partial S / \partial
r)$; therefore, one should perform the following change of the radial variable
\begin{eqnarray}
r \;
\Longrightarrow \;  r^{*}  \; , \quad { d \over dr } = {1 \over 1 -
{r^{2} / \rho^{2}}} {d \over dr^{*}}\;, \quad
r^{*} =  { \rho \over 2 } \ln  {1 +  r / \rho  \over 1 - r/ \rho }\; .
\label{3.6a}
\end{eqnarray}

\noindent The result is
\begin{eqnarray}
p_{r^{*}}  = {d \over dr^{*}}S = \left(1 - {r^{2} \over  \rho^{2}} \right) {d
S \over d r} =
                    \left(1 - {r^{2} \over  \rho^{2}} \right) p_{r} \;
\; \Longrightarrow \;\; p_{r^{*}}  \rightarrow  {(\epsilon +
e^{2} / \rho)^{2} \over c^{2}} \; .
\label{beh-3}
\end{eqnarray}

\noindent The corresponding plot of \ $p^{2}_{r^{*}}$ \ is shown in Fig. 3.
This picture is well understood from the physical point of view as one representing
a nonstable quantum-mechanical hydrogen atom model in the de Sitter space.
We note that one may expect a similar behavior when considering the atom model
on the basis of the Dirac equation.

\vspace{10mm}

\unitlength=0.6mm
\begin{picture}(100,80)(-60,-40)
\special{em:linewidth 0.4pt} \linethickness{0.4pt}

\put(0,0){\vector(+1,0){150}}   \put(150,-5){$r^{*}$}
\put(0,-40){\vector(0,+1){80}}  \put(+2,+40){$p_{r^{*}}^{2}$}

\put(+10,0){\line(-1,-4){5}} \put(+5,-40){\line(0,+1){20}}

\put(+10,0){\line(+1,+1){20}} \put(+30,+20){\line(+1,-1){20}}

\put(+50,+0){\line(+2,-1){20}} \put(+70,-10){\line(+1,0){20}}
\put(+90,-10){\line(+1,0){10}} \put(+100,-10){\line(+1,+1){10}}

\put(+110,0){\line(+2,+1){10}} \put(+120,+5){\line(+1,0){20}}

\end{picture}

\begin{center}

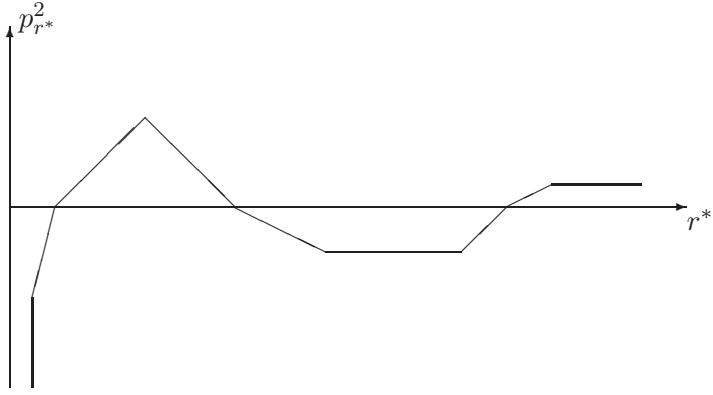
{\bf Fig.  3. Coulomb problem in $dS$ space, the coordinate $r^{*}$}.

\end{center}

\section{Reducing the quantum-mechanical radial equation to the general Heun equation}

The radial equation (\ref{2.3}) is a Fuchsian ordinary linear differential equation
having four regular singular points.
Using the dimensionless quantities
\begin{eqnarray}
x = { r \over  \rho } \; , \qquad {\epsilon \rho  \over c \hbar }
= E\;, \qquad {e^{2} \over  c \hbar } = \alpha \;, \qquad { M^{2}
c^{2} \rho^{2}   \over \hbar^{2} } \Longrightarrow M^{2}\; ,
\label{5.1}
\end{eqnarray}

\noindent and making the standard substitution $f=x^{A}\,(1-x)^{B}\,(1+x)^{C}\,H(x)$,
it is reduced to the general Heun equation
for $H(x)=H(-1,q;\lambda,\beta,\gamma,\delta;x)$ written in its canonical form as \cite{Ronveaux, Slavyanov-Lay}
\begin{eqnarray}
{d^{2}H\over dx^{2}}+ \left ( {\gamma\over x} + {\delta\over x-1} +
{\epsilon \over x+1}\right ) \,{dH\over d x} +
{\lambda \beta x-q\over  x(x-1)(x+1)}\,H=0 \; ,
\label{generalHeun}
\end{eqnarray}

\noindent where
\begin{eqnarray}
\gamma=2+2A\, , \; \delta=1+2B\, , \; \epsilon = 1+ 2 C  , \;
\label{parameter-1}
\\
\lambda,\beta ={3\over 2}\,+A+B+C  \pm  \sqrt{  - M^{2} + {9 \over 4}} \, ,
\label{parameter-2}
\\
q=2\,(E \alpha-(1+A)(B-C))  \,
\label{parameter-3}
\end{eqnarray}

\noindent with parameters $A,B,C$ given as
\begin{eqnarray}
A=-{1\over 2}\pm\sqrt{ (l+1/2)^{2} - \alpha^{2}}\,, \qquad B=\pm
{i\over 2}\,(E+\alpha)\, , \qquad C=\pm {i\over 2}\,(E-\alpha)\; .
\label{restriction-1}
\end{eqnarray}

\noindent Note that the singular point $x= +1$  of the Heun equation (\ref{generalHeun}) is physical,
because it represents the de Sitter horizon,
while the point  $x=-1$ does not belong to the physical domain, the interval  $x \in [0, +1 )$.

To get solutions vanishing at the origin, we take
\begin{eqnarray}
A=-{1\over 2} + \sqrt{ (l+1/2)^{2} - \alpha^{2}} \; . \label{5.7a}
\end{eqnarray}

\noindent Depending on the signs of $B$ and $C$, we have four choices, $(+,+)$, $(-,-)$, $(+,-)$ and $(-,+)$,
each leading to a different particular solution.
The parameters $\lambda,\beta$ and $q$ for these solutions read:
\begin{eqnarray}
(+,+)\qquad \lambda, \beta  ={3\over 2}\,+A +iE  \pm   i  \sqrt{  M^{2} -
{9 \over 4} }\; , \qquad q=  2 \alpha\;  [ \; E  -  i  (A+1) \; ] \,,
\nonumber
\\
(-,-)\qquad \lambda, \beta  ={3\over 2}\,+A -iE  \pm   i \sqrt{  M^{2} -
{9 \over 4} }\; , \qquad  q= 2   \alpha \;  [\;E  +  i (A+1) \; ] \,,
\nonumber
\\
(+,-)\qquad \lambda, \beta  ={3\over 2}\,+A + i \alpha  \pm i    \sqrt{
M^{2} - {9 \over 4} }\; , \qquad  q= 2 E \; [ \;  \alpha - i  (A+1) \; ] \,,
\nonumber
\\
(-,+)\qquad \lambda, \beta  ={3\over 2}\,+A -i \alpha  \pm   i  \sqrt{
M^{2} - {9 \over 4} }\; , \qquad q= 2 E  \; [ \;  \alpha + i   (A+1) \; ] \, .
\label{5.10b}
\end{eqnarray}

\noindent Note that by physical arguments we may assume the inequality
$
M^{2} \rightarrow  (M^{2} \rho^{2} c^{2} /  \hbar^{2} ) >> 1
$.

We conclude this section by discussing a speculative possibility of getting complex values for energy. This situation is the case if one assumes a possible  quantization rule  $\lambda = - n$, which may provide polynomial (thus, finite on the interval  $x \in [0, +1]$) solutions of Eq. (\ref{generalHeun}). This results in complex energies for the cases $(+,+)$ and $(-,-)$:
\begin{eqnarray}
(+,+)\;\;  E= +i\, \left({3\over 2}+n+A \right) -
\sqrt{M^{2}-{9\over 4}}\, , \qquad
(-,-)\;\;  E= -i\, \left({3\over 2}+n+A \right) + \sqrt{M^{2}-{9\over
4}}\, , \label{5.11}
\end{eqnarray}
so that (let $\Gamma = \sqrt{M^{2}- 9/  4}$)
\begin{eqnarray}
(+,+) \; e^{-i\epsilon t / \hbar} =   e^{     (ct / \rho ) [ +\,( 3/ 2+n+A ) + i \Gamma
] },\quad
(-,-) \; e^{-i\epsilon t / \hbar} =
 e^{     (ct / \rho ) [ -( 3/ 2+n+A ) - i  \Gamma ] } \; .
\label{5.12}
\end{eqnarray}
We note that these factors behave differently at $t \rightarrow +\infty $.
It seems that more physical is the variant $(-,-)$, since in this case
\begin{eqnarray}
(-,-) \quad  \left \{  (e^{-i\epsilon t / \hbar} )^{*}  e^{-i\epsilon t / \hbar} \right \} =
e^{   -  2(ct / \rho )  ( 3/ 2+n+A )  } \rightarrow 0 \;\; \mbox{when} \;\; t \rightarrow  + \infty \;.
\label{7.13}
\end{eqnarray}

The question concerning the physical meaning of complex values of the energy is open. The influence of the Coulomb potential on this complex-valued spectrum is little (see Eq. (\ref{5.7a}) for $A$), and the formulas (\ref{5.12}) have no correlation with the known spectrum in the flat Minkowski space.

\section{WKB-analysis}

To study the problem within the WKB approximation, we assume large $\rho$
and consider the region $r \ll \rho $, where the potential well is located.
Expanding the roots $r_{1},...,r_{4}$ of Eq. (\ref{2.5})
in terms of the small parameter $\rho^{-1}$, we get
$$
r_{i} = r_{0i} + {\Delta_{i} \over  \rho^{2} } \; , \qquad i =1,2 \; ,
$$
\begin{eqnarray}
r_{01,02} = { c^{2} \over  M^{2}c^{4}- \epsilon^{2} } \; \left  [ \;
{e^{2} \epsilon \over c^{2}} \pm M \sqrt{{e^{4}\epsilon^{2} \over
c^{4}}  - \left(L^{2} - {e^{4}\over c^{2}}\right) { M^{2} c^{2}-\epsilon^{2}
\over c^{2}}} \;  \; \right ]> 0  \; ,
\nonumber
\\
\Delta_{i} = - r_{0i}^{2} \; (L^{2} + M^{2} c^{2} r_{0i}^{2} ) \;
\left  [\;  {2 e^{2} \epsilon \over c^{2}} + 2 r_{0i}
\left({\epsilon^{2} \over c^{2}} - M^{2}c^{2}\right) \;\right ]^{-1}
\label{6.3c}
\end{eqnarray}

\noindent and
\begin{eqnarray}
r_{3,4} = \pm \rho \sqrt{1 - {\epsilon^{2} \over M^{2} c^{4}}}
 -{e^{2}\epsilon \over M^{2} c^{4} - \epsilon^{2}} \; , \qquad i =3,4 \; .
\label{6.4b}
\end{eqnarray}

It is immediately seen that at large curvature radius $\rho$
the region between $r_{1}$ and $r_{2}$ lays far from $r_{3}>0$ and  $ r_{4}<0$.
Then, rewriting
\begin{eqnarray}
p_{r}^{2} = { 1 \over r^{2} \left( 1 - {r^{2}/\rho^{2}} \right)^{2}} \;
{ M^{2} c^{2} \over \rho^{2}} \; (r-r_{1}) (r-r_{2})(r-r_{3})
(r-r_{4}) \;
\label{7.1}
\end{eqnarray}

\noindent and expanding the product $(r-r_{3})(r-r_{4})$ for $r \ll r_{3,4}$,
in virtue of Eq. (\ref{6.4b}), we obtain the approximate form for $p_{r}$
in the region $r \ll \rho $, where the potential well is located:
\begin{eqnarray}
p_{r} \sim M c {\sqrt{(r-r_{1})((r - r_{2}) } \over r }\left  (
{\epsilon^{2} \over M^{2} c^{4}} - 1 \right  )^{1/2} \left (
 1 + A \; {r^{2} \over \rho^{2}} \right ) , \;\; A = 1 - {1 \over 2 ( 1 -\epsilon^{2} /M ^{2} c^{4} ) }\;.
\label{7.2}
\end{eqnarray}

\noindent The WKB quantization rule is
\begin{eqnarray}
\int _{L} p_{r} dr = 2 \pi \hbar ( n+1/2) \; . \label{6.3}
\end{eqnarray}

\noindent  From this, taking  $p_{r}$  according to  (\ref{7.2}) and choosing the path of integration surrounding the turning points  $r_{1}$ and  $r_{2}$, we reduce the problem to calculating the residues just in two points, $r=0$ and $r=\infty$. As a result, we arrive at the equation

\begin{eqnarray}
\sqrt{ {\epsilon^{2} \over c^{2}} - M^{2} c^{2}} \;\left  [ \; \left(
\; \sqrt{r_{1} r_{2}}  + {r_{1} + r_{2} \over 2} \; \right)\; - \; {A
\over \rho ^{2}} r_{1} r_{2} {r_{1} + r_{2} \over 2} \;\right  ] =
i \hbar ( n + 1/2) \; ,
\label{7.5}
\end{eqnarray}

%
%

\noindent from where we get the following approximate formula for energy levels in the de Sitter space:
\begin{eqnarray}
\epsilon = \epsilon_{0}  + { \Delta \over \rho^{2} } \; , \;\;
\epsilon_{0} = Mc^{2}\;  \left [ \;  1 + {e^{4} /c^{2} \over [
\hbar (n+1/2) + \sqrt{ \hbar^{2}  (l+1/2)^{2} - e^{4} / c^{2} }
]^{2} } \;\right  ] ^{-1/2} \;
\label{7.8}
\end{eqnarray}
with
\begin{eqnarray}
\Delta = {(M^{2} c^{4} - \epsilon^{2}_{0} ) ^{2} \over 4 e^{2}
M^{2}c^{4} } \; \; \left [ \; \Delta_{1} + \Delta_{2}  +
\Delta_{1} \sqrt{{r_{02} \over r_{01}} } + \Delta_{2}  \sqrt{
{r_{01} \over r_{02} }}  - A r_{01} r_{02}(r_{01} + r_{02})
 \;  \right ]  ,
\label{Delta}
\end{eqnarray}
where $\Delta_{1,2}$, $r_{01,02}$ and $A$ are given by above
formulas with $\epsilon$ replaced by  $\epsilon_{0}$.

In accordance with the behavior of the function $p_{r}^{2}$, the particle
can move from the domain between $r_{1}$ and $r_{2}$
to the domain near the horizon $r \sim \rho $ due to the quantum-mechanical tunneling.
The probability of this process is given as
\begin{eqnarray}
W = \mbox{exp}  \;\left ( -{2 \over \hbar} \int_{r_{2}}^{r_{3}}
\sqrt{-p_{r}^{2}} \; dr  \right ) , \label{7.9}
\end{eqnarray}

\noindent from which we obtain a rough  estimation  $W \sim e^{-2\rho /
\lambda }$, where  $\lambda$ is the Compton wavelength.

\section{The hydrogen atom in $AdS$ space}

We now consider the Coulomb problem in static coordinates of the the anti de Sitter space \cite{Hawking-Ellis-1973}
\begin{eqnarray}
dS^{2} = \left (1+ {r^{2} \over \rho^{2}} \right ) dt^{2}  - \left (1+ {r^{2} \over
\rho^{2}} \right  )^{-1} dr^{2} - r^{2} ( d \theta^{2}  + \sin^{2} \theta
d \phi^{2} ) , \quad  r \in [0, + \infty )\; .
\label{8.1}
\end{eqnarray}

\noindent  From the Klein-Gordon-Fock equation, for the 4-potential
$ A_{\alpha} = ( {e/r}, 0,0,0) $  we get the radial equation (compare with Eq. (\ref{2.3}))
\begin{eqnarray}
{ d^{2} \over d r^{2} } \; f\;  +        \; {2 (1+2r^{2} /
\rho^{2}) \over r (1+ r^{2} / \rho^{2}) } \; {d \over dr } \; f \;
+ \qquad \qquad \quad
\nonumber
\\
+ \left [ { ( \epsilon + e^{2} / r ) ^{2} \over c^{2} \hbar ^{2} }
{ 1 \over (1 + r^{2} / \rho^{2} )^{2} } - \left ( { M^{2} c^{2}  \over
\hbar^{2} } + { l(l+1) \over r^{2} } \right ) {1 \over 1 + r^{2} /
\rho^{2} }\right  ] f = 0 \;.
\label{8.3}
\end{eqnarray}
In dimensionless form, using
\begin{eqnarray}
x = { ir \over  \rho } \; , \qquad {\epsilon \rho  \over c \hbar }
= E\;, \qquad {e^{2} \over  c \hbar } = \alpha \;, \qquad { M^{2}
c^{2} \rho^{2}   \over \hbar^{2} } \Longrightarrow M^{2}\,,
\end{eqnarray}

\noindent and
applying the substitution
$f=x^{A}\,(x-1)^{B}\,(x+1)^{C}\,H$
with parameters
\begin{eqnarray}
A=-{1\over 2}\pm\sqrt{ (l+1/2)^{2} - \alpha^{2}}\,, \qquad B=\pm
{1\over 2}\,(E+i \alpha)\,,\qquad C=\pm {1\over 2}\,(E-i \alpha)\; ,
\label{restriction-2}
\end{eqnarray}

\noindent we reduce Eq. (\ref{8.3}) to the general Heun equation for the function
 $H(-1,q;\lambda,\beta,\gamma,\delta,x)$:
\begin{eqnarray}
{d^{2}H\over dx^{2}}+ \left ( {\gamma\over x} + {\delta\over x-1} +
{\epsilon \over x+1}\right ) \,{dH\over d x} +
{\lambda \beta x-q\over  x(x-1)(x+1)}\,H=0 \; , \label{8.8b}
\end{eqnarray}

\noindent the parameters of which are given as
\begin{eqnarray}
 \gamma=2+2A\, , \; \delta=1+2B\, , \; \epsilon = 1+ 2 C  , \;
\nonumber
\\
\lambda,\beta={3\over 2}\,+A+B+C  \pm  \sqrt{ M^{2} + {9 \over 4}}\, ,
\nonumber
\\
q=-2\,(i E \alpha+(1+A)(B-C))\, ,
\label{parameter-2}
\end{eqnarray}
\noindent where the upper sign $(+)$ conventionally stands for $\lambda$
and the lower sign $(-)$ refers to $\beta$.
Note that the variation region for the radial variable here is the interval
$x \in [ 0, +i \infty )$.
To have finite solutions at the origin, we should use the positive value of $A$:
\begin{eqnarray}
A=-{1\over 2} + \sqrt{ (l+1/2)^{2} - \alpha^{2}}> 0 \; .
\label{8.10}
\end{eqnarray}

\noindent Depending on the signs of $B$ and $C$,
we have four different possibilities: $(+,+)$, $(-,-)$, $(+,-)$ and $(-,+)$,
each providing a different particular solution:
\begin{eqnarray}
(-,-):\; \qquad \lambda = 3 /  2\,+A-E  +  \sqrt{  M^{2} + 9 /  4 }\; ,
\quad q = -2  i\alpha \; [ \;  E -  (A+1) \; ] \; ,
\nonumber
\\
(+,+):\; \qquad \lambda = 3 / 2 \,+A+E  + \sqrt{  M^{2} + 9/  4 }\; ,
\quad q = -2  i\alpha\;  [ \; E +  (A+1)\; ] \; ,
\nonumber
\\
(+,-):\; \qquad \lambda  = 3 /  2 \,+A+ i \alpha  +  \sqrt{  M^{2}+ 9 /  4 }\; ,
\quad q = -2E \;  [ \; i\alpha  +   (A+1) \; ] \; ,
\nonumber
\\
(-,+):\; \qquad \lambda  = 3/ 2 \,+A-i\alpha  +      \sqrt{ M^{2} + 9 / 4 }\; ,
\quad q = -2 E \; [ \;  i\alpha  -   (A+1) \; ]\; .
\label{8.13}
\end{eqnarray}

To choose the appropriate branch, consider the situation without Coulomb field, $\alpha=0$.
The Heun equation (\ref{8.8b}), after transforming it to the variable $y=x^{2}$,
becomes a hypergeometric equation
\begin{eqnarray}
 { d^{2} f\over d y^{2} }   +
   \left (  \; {3/2\over y}+{1\over y-1}  \right  )\; {d f\over d y }
   +\left [  \;-{1\over 4}\,{l(l+1)\over y^{2}}-{1\over 4}\,{E^{2}\over (y-1)^{2}}+\right.
\nonumber
\\
\left.+{1\over 4}\,{E^{2}-M^{2}+l(l+1)\over y-1}+{1\over
4}\,{-E^{2}+M^{2}-l(l+1)\over y}\right  ]f = 0 \;. \label{8.17}
\end{eqnarray}

\noindent Using the substitution
$f=y^{A}(y-1)^{B}\;_2F_1(a,b;c;y)$ with
$
A=+ l/ 2\,,\; B= -E / 2 $,
we construct the solution in terms of the hypergeometric function
with the parameters
\begin{eqnarray}
a = {1 \over  2} \left ( {3 \over 2} +l
- E   +  \sqrt{9/4 + M^{2}}\; \right ) \,, \;\;
b = {1 \over  2} \left ( {3 \over 2} +l
- E   -  \sqrt{9/4 + M^{2}} \right ) \,, \;\;
c =l + {3\over 2}\, .
\label{8.20c}
\end{eqnarray}
The hypergeometric function reduces to polynomials if $a=-n$ or $b=-n$.
The permissible choice is $a=-n$ since then the complete radial function
will vanish at infinity owing to the inequality
\begin{eqnarray}
A+B+n = {1\over 2} \left ( - {3 \over 2}  - \sqrt{9/4 + M^{2}} \right )  <0 \; .
\label{inequality}
\end{eqnarray}
The result is the discrete energy spectrum given as
\begin{eqnarray}
 E_{n} = 2n +l +{3\over 2} +  \sqrt{{9 \over 4} +M^{2}}\; , \qquad n = 0, 1, 2, ... \; .
\label{8.22}
\end{eqnarray}
Inspecting now Eqs. (\ref{8.13}), we find that this result is recovered, by applying the quantization rule in the form  $\lambda = -2n$ with the parameter $\lambda$ determined by the first line. Thus, the appropriate choice for the signs of $A$ and $B$ is $(-,-)$.

\section{Qualitative study of the problem in $AdS$ space}

Consider the locations of the turning points of the classical momentum.
By Vieta's theorem, the roots of the equation $p^{2}_{r}=0$ obey the identities
\begin{eqnarray}
r_{1}+r_{2}+r_{3}+r_{4}=0\, ,
\label{9.10a}
\\
r_{1}r_{2}r_{3}r_{4}= + {(L^{2} - e^{4} /c^{2})\rho^{2}
\over {M^{2}c^{2}}} >0 \; .
\label{r1r2r3r4'}
\end{eqnarray}

\noindent The last relation shows that, assuming all four roots real-valued, the variants
\begin{eqnarray}
\textrm{sign}(r_{1}, r_{2}, r_{3}, r_{4}) = (-,-,-, +)\; , \qquad
\textrm{sign}(r_{1}, r_{2}, r_{3}, r_{4}) = (+,+,+, -)\; .
\end{eqnarray}
\noindent are forbidden. Besides, the variants
\begin{eqnarray}
\textrm{sign}(r_{1}, r_{2}, r_{3}, r_{4}) = (-,-,-, -)\; ,\qquad
\textrm{sign}(r_{1}, r_{2}, r_{3}, r_{4}) = (+,+,+, +)\;
\end{eqnarray}

\noindent are also forbidden because of the identity (\ref{9.10a}). So, we conclude that there exists only one possibility for the real roots' location:
\begin{eqnarray}
\textrm{sign}(r_{1}, r_{2}, r_{3}, r_{4}) = (-,-,+, +)\; , \qquad
r_{4} > r_{3} > 0 \; ,
\label{9.11}
\end{eqnarray}

\noindent where   $r_{4} ,   r_{3}  $ stand for physical turning points.
Note also that the roots  $r_{1},\;r_{2}$  can be complex and conjugate:
\begin{eqnarray}
\textrm{sign}(r_{1}, r_{2}, r_{3}, r_{4}) \sim (z, z^{*}, +,+,)\; , \quad
\mbox{Re}\; z < 0 \; .
\label{9.16}
\end{eqnarray}

Thus, the qualitative study shows that in $AdS$ space we can expect for the Coulomb problem the situation of a complicated potential well with normal (real) energy spectrum, so that the hydrogen atom is expected to be a stable system in the quantum-mechanical sense.

\section{WKB-treatment for $AdS$ space}

We use the formal method of analytical continuation, which gives the possibility to get results without repeating much the same calculations:
\begin{eqnarray}
dS \Longrightarrow AdS\qquad
\rho \Longrightarrow i\rho \; .
\label{10.1'}
\end{eqnarray}

The approximate roots $r_{1}, r_{2}$ of the equation $p^{2}(r)=0$ in this case read
\begin{eqnarray}
r_{i} = r_{0i} - {\Delta_{i} \over \rho^{2}}\; , \qquad i =1,2 \; , \hspace{30mm}
\nonumber
\\
r_{01,02} = { c^{2} \over  M^{2}c^{4}- \epsilon^{2} }  \left  [
{e^{2} \epsilon \over c^{2}} \pm M \sqrt{{e^{4}\epsilon^{2} \over
c^{4}}  - \left( L^{2} - {e^{4}\over c^{2}} \right) { M^{2} c^{2}-\epsilon^{2}
\over c^{2}}}  \right ]> 0  \; ,
\nonumber
\\
\Delta_{i} = - r_{0i}^{2} \; (L^{2} + M^{2} c^{2} r_{0i}^{2} ) \;
\left  [  {2 e^{2} \epsilon \over c^{2}} + 2 r_{0i}
\left({\epsilon^{2} \over c^{2}} - M^{2}c^{2} \right) \right ]^{-1}\; ,
\label{10.2'}
\end{eqnarray}

\noindent and for the roots $r_{3}, r_{4}$ we have
\begin{eqnarray}
r_{i} = i\rho \left( b_{0i} + {b_{1} \over i\rho} \right) \; , \quad i=3,4 \; , \quad
\nonumber
\\
b_{03,04} = \pm \sqrt{1- {\epsilon^{2} \over M^{2} c^{4}}} \; , \quad
b_{1} = - {e^{2}\epsilon \over M^{2} c^{4} - \epsilon^{2}} \; .
\label{10.3'}
\end{eqnarray}

Accordingly, the approximate form of $p_{r}$ in the region $r \ll \rho $
(where the potential well is located) is written as
\begin{eqnarray}
p_{r} \sim M c {\sqrt{(r-r_{1})((r - r_{2}) } \over r } \left(
{\epsilon^{2} \over m^{2} c^{4}} - 1  \right)^{1/2}\; \left (
 1  -  A \; {r^{2} \over \rho^{2}} \right )\; ,\; \;
 A = 1 - {1 \over 2 ( 1 -\epsilon^{2} / M^{2} c^{4} ) } \; .
\label{10.4'}
\end{eqnarray}

The quantization condition is the same as Eq. (\ref{6.3}). Choosing the contour of integration surrounding the points $r_{1}$ and $r_{2}$, we again reduce the problem to finding the residues at $r=0$ and $r=\infty$
and arrive at the equation
\begin{eqnarray}
\sqrt{ {\epsilon^{2} \over c^{2}} - M^{2} c^{2}} \left  [ \left  (
\sqrt{r_{1} r_{2}}  + {r_{1} + r_{2} \over 2} \right  ) +  {A \over
\rho ^{2}} r_{1} r_{2} {r_{1} + r_{2} \over 2} \right  ] = i \hbar ( n
+ 1/2) \; .
\label{quant-2}
\end{eqnarray}

\noindent From this, an approximation for the energy spectrum in $AdS$ space follows:
\begin{eqnarray}
\epsilon = \epsilon_{0}   - { \Delta \over \rho^{2} } \; ,
\quad
\epsilon_{0} = Mc^{2}  \left [   1 + {e^{4} /c^{2} \over [ \hbar
(n+1/2) + \sqrt{ \hbar^{2}  (l+1/2)^{2} - e^{4} / c^{2} }   ]^{2} }
\right  ] ^{-1/2},
\nonumber
\\
\Delta = {(M^{2} c^{4} - \epsilon^{2}_{0} ) ^{2} \over 4 e^{2}
M^{2}c^{4} }   \left [  \Delta_{1} + \Delta_{2}  +
\Delta_{1} \sqrt{{r_{02} \over r_{01}} } + \Delta_{2}  \sqrt{
{r_{01} \over r_{02} }}   + A r_{01} r_{02}(r_{01} + r_{02})
   \right ] ,
\label{10.6'}
\end{eqnarray}
where $\Delta_{1,2}$, $r_{01,02}$ and $A$ are given by above
formulas with $\epsilon$ replaced by  $\epsilon_{0}$.

\section{Spin 1/2 particle in the Coulomb field in $dS$ and $AdS$ spaces }

To discuss the case of a spin 1/2 particle, we start with the Dirac free wave equation (the notations of \cite{Book-2009} are used)
\begin{eqnarray}
\left [\; i \gamma^{c} \; \left ( e_{(c)}^{\alpha} \partial_{\alpha} +
 {1 \over 2} \sigma^{ab} \gamma_{abc} \right ) -  M \; \right  ]
 \; \Psi = 0 \; .
\label{Dirac}
\end{eqnarray}

\noindent
In static coordinates of the de Sitter space, using the spinor basis and dimensionless coordinate $r \in [0, 1)$, applying the substitution
$ \Psi  (x)  = r ^{-1} \Phi ^{-1/4} \; F (x)
 $, $\Phi = 1 -r^{2}$,
 this equation is reduced to the form \cite{Book-2011}
 \begin{eqnarray}
 \left [ \;i\; {\gamma ^{0}  \over \sqrt{\Phi} }
\partial _{t} \; + i \sqrt{\Phi } \gamma^{3}\;   \;
\partial _{r}  +   { 1 \over r} \;\Sigma _{\theta,\phi }  -  M \;
  \right ]  \; F (x)  =  0  \; ,
\label{11.0}
\\
  \Sigma _{\theta ,\phi } =  \; i\; \gamma ^{1}
\partial _{\theta } + \gamma ^{2} {i\partial +i\sigma^{12}\cos
\theta  \over \sin \theta  }\; . \quad \quad
\label{11.1}
\end{eqnarray}

\noindent Spherical waves are constructed as
\begin{eqnarray}
\Psi _{\epsilon jm}(x) \;  = \; {e^{-i\epsilon t} \over r} \;
\left | \begin{array}{l}
        f_{1}(r) \; D_{-1/2} \\ f_{2}(r) \; D_{+1/2}  \\
        f_{3}(r) \; D_{-1/2} \\ f_{4}(r) \; D_{+1/2}
\end{array} \right | \; ,
\label{11.2}
\end{eqnarray}

\noindent where the Wigner functions are denoted according to the rule
$D^{j}_{-m, \sigma}(\phi, \theta, 0) =D_{\sigma}$. With the use of the recurrence relations \cite{Varshalovich-Moskalev-Hersonskiy-1975}
\begin{eqnarray}
\partial_{\theta} \; D_{+1/2} \; = \;  a\; D_{-1/2}  - b \; D_{+3/2}  \; ,
\;
{- m - 1/2 \;  \cos \theta  \over  \sin \theta } \; D_{+1/2}\; =\;
- a \;  D_{-1/2} - b \;  D_{+3/2}  \; ,
\\
\partial_{\theta} \;  D_{-1/2} \; = \;  b \; D_{-3/2} - a \; D_{+1/2}  \; ,
\;
{- m + 1/2 \; \cos \theta \over \sin \theta}  \;  D_{-1/2} \; = \;
 - b \; D_{-3/2}  - a \;  D_{+1/2}  \; ,
\end{eqnarray}

\noindent where
$
a = (j + 1)/2\;, \; b  = (1 / 2) \;\sqrt{(j-1/2)(j+3/2)}\;,
$
we get
\begin{eqnarray}
 \Sigma _{\theta ,\phi } \; \Psi _{\epsilon jm}(x) \;  = \; i\; \nu \;
{e^{-i\epsilon t } \over r} \; \left | \begin{array}{r}
        - \; f_{4}(r) \; D_{-1/2}  \\  + \; f_{3}(r) \; D_{+1/2} \\
        + \; f_{2}(r) \; D_{-1/2}  \\  - \; f_{1}(r) \; D_{+1/2}
\end{array} \right |    ,
\label{11.3}
\end{eqnarray}

\noindent where $\nu  = (j + 1/2)$. This leads to the following radial system:
\begin{eqnarray}
{\epsilon  \over \sqrt{\Phi}}   f_{3}   -  i  \sqrt{\Phi} {d \over dr}  f_{3}   - i {\nu \over r}
f_{4}  -  M  f_{1} =   0  \; ,
\;\;
{\epsilon  \over \sqrt{\Phi}}   f_{4}   +  i \sqrt{\Phi}  {d
\over dr} f_{4}   + i {\nu \over r} f_{3}  -  M  f_{2} =   0  \; ,
\label{11.444}
\\
{\epsilon  \over \sqrt{\Phi}}   f_{1}   +  i \sqrt{\Phi}  {d \over dr}  f_{1}  + i {\nu \over r}
f_{2}  -  M  f_{3} =   0  \; ,
\;\;
{\epsilon   \over \sqrt{\Phi}} f_{2}   -  i\sqrt{\Phi}   {d
\over dr} f_{2}   - i {\nu \over r} f_{1}  -  M  f_{4} =   0  \; .
\label{11.4}
\end{eqnarray}

\noindent To simplify the system, we diagonalize the $P$-operator
\cite{Book-2011}. In the Cartesian basis, $\hat{\Pi}_{C} = i
\gamma ^{0} \otimes \hat{P}$, the transition to spherical tetrad gives
\begin{eqnarray}
\hat{\Pi}_{sph} \; \; = \left | \begin{array}{cccc}
0 &  0 &  0 & -1   \\
0 &  0 & -1 &  0   \\
0 &  -1&  0 &  0   \\
-1&  0 &  0 &  0
\end{array} \right |
\; \otimes  \; \hat{P} \; .
\label{11.5}
\end{eqnarray}

\noindent From the equation
$\hat{\Pi}_{sph} \; \Psi _{jm} = \; \Pi \; \Psi _{jm}$ it follows that
$\Pi =  \delta \;  (-1)^{j+1} , \; \delta  = \pm 1 $ and
\begin{eqnarray}
f_{4} = \; \delta \;  f_{1} , \;  f_{3} = \;\delta \; f_{2} \;
,\;
\Psi (x)_{\epsilon jm\delta } \; = \; {e^{-i\epsilon t} \over  r }
\; \left | \begin{array}{r}
     f_{1}(r) \; D_{-1/2} \\
     f_{2}(r) \; D_{+1/2} \\
\delta \; f_{2}(r) \; D_{-1/2}   \\
\delta \; f_{1}(r) \; D_{+1/2}
\end{array} \right |  \; .
\label{11.10}
\end{eqnarray}

\noindent The system (\ref{11.444})-\ref{11.4}) is then simplified as
\begin{eqnarray}
\left( \sqrt{\Phi} {d \over dr} + {\nu \over r} \right)  f  +  \left( { \epsilon  \over \sqrt{\Phi}}  +
 \delta  M \right) g  = 0 \; ,\quad
\left( \sqrt{\Phi} {d \over dr} - {\nu \over r} \right) g  -  \left( {\epsilon  \over \sqrt{\Phi}} -
 \delta  M \right) f = 0     \; ,
\label{11.11}
\end{eqnarray}

\noindent where the new functions
$
f  = ( f_{1} + f_{2} )/ \sqrt{2} \; , \;  g  =
(f_{1} - f_{2} ) /  i \sqrt{2}$
are used instead of $f_{1}$  and  $f_{2}$.
For definiteness, let us consider Eqs. (\ref{11.11}) for  $\delta = +1$
(formally, the second case $\delta =-1$ corresponds to the change $M \rightarrow - M$):
\begin{eqnarray}
\left( \sqrt{\Phi} {d \over dr} + {\nu \over r} \right)  f  +  \left( { \epsilon  \over \sqrt{\Phi}}  +
  M \right) g  = 0 \; ,\;
\left( \sqrt{\Phi} {d \over dr}  - {\nu \over r} \right) g  -  \left( {\epsilon  \over \sqrt{\Phi}} -
  M \right) f = 0     \; .
\label{11.12}
\end{eqnarray}

To take into account the Coulomb field, it suffices to make the formal change $\epsilon \; \rightarrow \epsilon +{e^{2}/r}$. In this way we get the system
\begin{eqnarray}
\left ( \sqrt{1-r^{2}} {d \over dr} + {\nu \over r} \right )  f  +
\left ( { 1  \over \sqrt{1 -r^{2} }} \left(\epsilon +{e^{2} \over r}\right)  +
  M  \right ) g  = 0 \; ,
\label{11.1333}
 \\
\left ( \sqrt{1-r^{2}}  {d \over dr}  - {\nu \over r} \right ) g  -
\left ( {1 \over \sqrt{1 - r^{2}}  } \left(\epsilon +{e^{2} \over r}\right) -
  M \right ) f = 0     \; .
\label{11.13}
\end{eqnarray}

\noindent Using the variable $\rho$ ($\sin\rho=r)$, these equations become shorter:
\begin{eqnarray}
\left ( {d \over d \rho} +  {\nu \over \sin \rho} \right )  f  +
\left  ( { 1 \over \cos \rho }\left(\epsilon +{e^{2} \over \sin \rho}\right)
+
  M  \right ) g  = 0 \; ,
\label{11.14a}
\\
\left (  {d \over d \rho}  - {\nu \over \sin \rho} \right ) g  -
\left ( {1 \over \cos \rho } \left(\epsilon +{e^{2} \over \sin
\rho }\right)  -  M \right )  f =  0     \; .
\label{11.14b}
\end{eqnarray}

\noindent Introducing two new functions:
\begin{eqnarray}
f + g = e^{-i\rho/2} (F + G) \; , \qquad
f - g = e^{+i\rho/2} (F - G) \; ,
\label{11.15}
\end{eqnarray}
\noindent and using a new variable $y=\tan(\rho/2)$ with the notation
\begin{eqnarray}
y_{1,2} = - { (\epsilon + M - i \nu - i /2) \pm \sqrt{(\epsilon + M - i \nu - i /2)^{2} - e^{4} }\over e^{2}}
\; ,
\label{11.21}
\\
Y_{1,2} = - { (\epsilon - M - i \nu + i /2) \pm  \sqrt{(\epsilon - M - i \nu + i /2)^{2} - e^{4} }\over e^{2}}
\; ,
\label{11.22}
\end{eqnarray}

\noindent we transform  Eqs. (\ref{11.14a}) and (\ref{11.14b}) to the form
\begin{eqnarray}
\left[(1+y^{2})\, {d\over dy}-  \nu y  + {\nu\over y} -
{ a \over 1-y}+{ b \over 1+y}  + {a-b \over 2}  \right]F+
 {e^{2} \over y} (y -y_{1}) (y- y_{2} )\; G=0\,,
\\
\left[ (1+y^{2})\,{d\over dy}+ \nu \,y  - {\nu\over y}+
{ a \over 1-y}-{ b \over 1+y}  - {a-b \over 2}  \right]G -
 {e^{2} \over y} (y -Y_{1}) (y- Y_{2}) \; F=0\,,
\label{11.23}
\end{eqnarray}

\noindent
where
$
a = 2i (\epsilon + e^{2}) \; , \; b = 2i( \epsilon - e^{2}) \; .
$
Elimination of $G$ (or $F$) from this system results in a second order differential equation for $F(y)$ (or $G(y)$), which has 8 singular points:  $0, \pm 1, \pm i,  y_{1,2}, \infty$ (or $0, \pm 1, \pm i,  Y_{1,2}, \infty$).

A similar situation is faced for the Dirac equation in the Coulomb potential on the background of the anti de Sitter space-time. The analysis here is much similar to that for the above de Sitter model. Using the variable
$ r = \sinh \rho  $,   instead  of Eqs. (\ref{11.14a}),(\ref{11.14b}) we have the equations
\begin{eqnarray}
\left ( {d \over d \rho} +  {\nu \over \sinh \rho} \right )
f + \left  ( { 1 \over \cosh \rho }(\epsilon +{e^{2} \over
\sinh \rho}) +
  M  \right ) g  = 0 \; ,
\\
\left (  {d \over d \rho}  - {\nu \over \sinh \rho} \right )
g - \left ( {1 \over \cosh \rho } (\epsilon +{e^{2} \over
\sinh \rho })  -  M \right )   f =  0     \; . \label{12.14}
\end{eqnarray}

\noindent Introducing now new functions $F$ and $G$:
\begin{eqnarray}
f +  g = e^{-\rho/2} ( F +  G) \; , \qquad
f - g = e^{+\rho/2} ( F -  G) \; ,
\label{12.15}
\end{eqnarray}

%



\noindent and a new variable $y=\tanh(\rho/2)$, this system is rewritten as
\begin{eqnarray}
\left[(1-y^{2})\, {d\over dy}+  \nu \;y  + {\nu\over y} -
{4(\epsilon y+e^{2})\over 1+y^{2}}  +2e^{2}  \right]F+
 {e^{2} \over y} (y -\bar{y}_{1}) (y- \bar{y}_{2} )\; G=0\,,
\\
\left[ (1-y^{2})\,{d\over dy}- \nu \;y  - {\nu\over y}+
{4(\epsilon y+e^{2})\over 1+y^{2}}  -2e^{2}  \right]G +
 {e^{2} \over y} (y -\bar{Y}_{1}) (y- \bar{Y}_{2}) \; F=0\,.
\label{12.23}
\end{eqnarray}

\noindent where
\begin{eqnarray}
\bar{y}_{1,2} =  { (\epsilon + M -  \nu - 1 /2) \pm
\sqrt{(\epsilon + M - \nu - 1 /2)^{2} + e^{4} }\over e^{2}} \; ,
\label{12.21}
\\
\bar{Y}_{1,2} =  { (\epsilon - M -  \nu + 1 /2) \pm
\sqrt{(\epsilon - M - \nu +1 /2)^{2} + e^{4} }\over e^{2}} \; .
\label{12.22}
\end{eqnarray}

Again, as in the case of the de Sitter space, the second order differential equations for $F(y)$ and $G(y)$ have eight singularities: $0, \pm 1, \pm i,  \bar{y}_{1}, \bar{y}_{2}, \infty$ and $0, \pm 1, \pm i,   \bar{Y}_{1}, \bar{Y}_{2}, \infty$, respectively.

\section{Summary}

Thus, we have developed the theory of the hydrogen atom in
 the de Sitter and anti de Sitter spaces on the basis of the Klein--Gordon--Fock wave equation in static coordinates. In both models, after separation of the variables, the problem is reduced to the general Heun equation, which is a second order Fuchsian linear differential equation having four regular singular points. The appearance of the Heun functions in treating the simplest and basic quantum mechanical system, the hydrogen atom, on the de Sitter geometrical backgrounds seems to be a useful point because of the progress in the theory of these functions observed during the last years. Conversely, the observed frequent appearance of the Heun equations in relativistic quantum-mechanical problems on the curved space-time background can serve as an additional stimulus to elaborate the mathematical theory of these functions.

A qualitative examination shows that the energy spectrum for the hydrogen atom
 in the de Sitter space is quasi-stationary, and the hydrogen atom is unstable in the sense
  of the quantum mechanics. We have derived an approximate expression for the energy
  levels within the quasi-classical method, and have estimated the probability
   of decay of the hydrogen atom. In contrast to this, the analysis shows that in
    the anti de Sitter model the hydrogen atom is stable. Approximate formulas for
    energy levels within the quasi-classical approach are also derived for this case.

Finally, we have extended the developments to the case of a spin 1/2 particle for both
de Sitter space models. This extension results in complicated second order differential
equations having 8 regular singular points.

We may expect (because of the universal character of the influence of the
gravitational forces) that the behavior of the hydrogen atom model is typical
 for all quantum mechanical systems in the de Sitter geometries: any system,
 being stationary in the Minkowski space, will become quasi-stationary on
  the background of the de Sitter geometry, whereas in the case of the anti
  de Sitter model it preserves the stability property in the quantum-mechanical sense.

Going beyond, we further conjecture that the influence of any fixed type of
 a space-time geometry is in a sense universal and is much the same for various
  quantum mechanical systems, irrespective of the spin of the involved particles
  (various examples supporting this supposition are presented in
  \cite{Book-2011, Book-2012, Book-2014}).

Finally, it should be mentioned that the de Sitter cosmological models can
be parameterized by non-static coordinates
\cite{Hawking-Ellis-1973}, in which the metrical coefficient $g_{00} =1$.
 For such cases, the Schr\"{o}dinger-type non-relativistic wave equations do
 exist, and one can develop a traditional quantum-mechanics influenced by a non-trivial
 geometry for the space-time in cosmological scales.

\section{Acknowledgements}
This work was supported by the Fund for Basic Researches of Belarus (Grant No. F14ARM-021) and by the Armenian State Committee of Science (Grant No. 13RB-052), within the cooperation framework between Belarus and Armenia.
The research has been partially conducted within the scope of the International Associated Laboratory (CNRS-France \& SCS-Armenia) IRMAS.
The work by A. Ishkhanyan has received funding from the European Union Seventh Framework Programme, grant No. 295025 - IPERA).

\end{document}